\documentclass{cargese}\usepackage{graphicx}

\let\footnote\savefootnote
\let\footnotetext\savefootnotetext 
\let\footnoterule\savefootnoterule 
\normallatexbib
\begin{document}

\articletitle[Moduli Stabilization from Fluxes]
{Moduli Stabilization from Fluxes}

\author{Michael B. Schulz}

\affil{California Institute of Technology 452-48\\
Pasadena, CA 91125 USA}    

\email{mschulz@theory.caltech.edu}

\begin{abstract}
Compactifications of IIB string theory with internal NS and RR
three-form flux are computationally attractive in that the lifting of
moduli is via a {\it perturbative\/} and often explicitly calculable
(super)potential.  We focus on the $T^6/Z_2$ orientifold, and provide
an illustative ${\cal N} = 2$ example.  For other choices of flux, the
resulting equations of motion can be solved to yield ${\cal N}=0,1,2$
or 3 supersymmetry in four dimensions.\hfill(arXiv:\,0810.5197,
CALT-68-2441)\par
\end{abstract}


Moduli are a familiar by-product of string compactifications, but they
do not exist in nature. Massless or nearly massless gravitationally
coupled scalars generate long range interactions that have been
excluded by fifth force experiments~\cite{Smith}.  They are also
problematic in cosmology~\cite{Linde}.\footnote{As the early universe
  cools, approximate moduli can easily overshoot the minima of their
  potentials. When this happens, they contribute an energy density
  like that of matter rather than radiation, causing deviations from
  the successful predictions of big bang nucleosynthesis.}

In traditional ${\cal N}=1$ heterotic compactifications,
perturbatively massless moduli can be lifted by well-known
nonperturbative effects like world-sheet instantons or Euclidean NS
five-brane instantons.  However, the computational difficulty of
quantitatively understanding these effects has so far proven
insurmountable.  There does not seem to exist a single compact example
in which anyone has computed the relevant instanton sums and
explicitly found a supersymmetric minimum of the resulting potential.

In contrast, when one turns on NS and RR three-form flux through
nontrivial three-cycles in a four-dimensional compactification of IIB
string theory, many of the moduli are lifted by a {\it perturbative\/}
scalar potential~\cite{GKP}
\begin{equation}
\label{potential}
{\cal V} \propto \int d^6y \sqrt{g_6}\, \bigl|G^{\rm ISD}\bigr|^2,
\quad G = F^{\rm RR}_{(3)} - \varphi H^{\rm NS}_{(3)}.
\end{equation}
Here, $\varphi$ is the dilaton-axion and $G^{\rm ISD}$ is the
imaginary-anti-self-dual part of the complex flux $G$.\footnote{Since
the compact manifold is six-dimensional, the hodge star operator
squares to $-1$ and its eigenvalues are $\pm i$.  The corresponding
eigenfunctions are {\it imaginary-self-dual} (ISD) three-forms
($*a_{(3)} = ia_{(3)}$) and {\it imaginary-anti-self-dual} (IASD)
three-forms ($*a_{(3)} = - ia_{(3)}$).}  This potential descends from
the compact part of the flux kinetic terms in the ten-dimensional IIB
supergravity action.

For consistency of the compactification, the fluxes must satisfy a
D3-brane charge tadpole cancellation condition, 
\begin{equation}
\label{tadpole}
N_{\rm D3} + \frac{1}{2(2\pi)^4\alpha'^2} \int H^{\rm NS}_{(3)}\wedge
F^{\rm RR}_{(3)} - \frac{1}{4} N_{\rm O3} = 0.
\end{equation}
in which wedged fluxes contribute in exactly the same way as
space-filling D3-branes and O3-planes.\footnote{In F-theory
compactifications, there would also be a contribution to
Eq.~(\ref{tadpole}) from the Euler character of the fourfold, and from
instantons on the compact part of D7-brane worldvolumes.}  An obvious
way to satisfy this condition is to start with a consistent string
compactification involving space-filling D3-branes, and then construct
new vacua by simply trading off D3-branes for fluxes.

However, whereas space-filling D3-branes and O3-planes preserve ${\cal
N}=4$ supersymmetry in four dimensions, the fluxes preserve less.  To
preserve at least ${\cal N}=1$ supersymmetry, we need vanishing
dilatino variation, and vanishing gravitino variation for at least one
gravitino.  This implies that $G$ must be primitive ({\it i.e.},
$J\wedge G = 0$, with $J$ the K\"ahler form), and of type
(2,1).\footnote{It can be shown that if $G$ is (2,1) and primitive
then it is also ISD.  So, if the supersymmetry conditions are
satisfied, then the scalar potential (\ref{potential}) is automaticaly
minimized and equal to zero.}

The (2,1) condition can be imposed by a superpotential~\cite{GVW}
\begin{equation}
\label{superpot}
W = \int G\wedge\Omega,
\end{equation}
where $\Omega$ is the holomorphic (3,0) form.  For proper Calabi-Yau
compactification, the primitivity condition is trivial due to the
absence of a fifth cohomology class. More generally, it is a linear
constraint on the K\"ahler moduli which is easy to solve.  On the
other hand, the (2,1) condition is not so simple.  When the periods of
the holomorphic (3,0) form are known, we can compute the
superpotential (\ref{superpot}).  But for Calabi-Yau orientifolds or
F-theory compactifications, this generally involves complicated
trancendental functions, for which it has not been possible to vary
the superpotential and solve the resulting equations, except near
singular (conifold) points in the moduli space of complex structure.

Still, one might expect that the equations of motion {\it are} soluble
when we choose a simple enough compactification manifold, and simplest
choice is a torus.  That is the choice we will make here, with one
modification.  Since we would like to turn on flux,
Eq.~(\ref{tadpole}) requires that there also be orientifold planes.
So, the compactification that we will consider is on the torus
orientifold $T^6/Z_2$~\cite{FP,KST}.

To define the $T^6/Z_2$ orientifold, we first compactify on $T^6$,
defined by $x^i\cong x^i+1$, $y^i\cong y^i+1$, $i=1,2,3$.  Then, we
mod out by the $Z_2$ parity operation $\Omega R_6(-1)^{F_L}.$ Here,
$\Omega$ is worldsheet parity, $R_6$ is a reflection of all of the
$T^6$ coordinates, and $(-1)^{F_L}$ is a parity operation that is
required by supersymmetry.\footnote{For massless modes, $(-1)^{F_L}$
acts as $-1$ on left-moving Ramond sector states and $+1$ otherwise.
If this factor were not included, the resulting spectrum of states
would not fill out supergravity multiplets.}  The massless states that
survive the orientifold projection are the four-dimensional graviton
$g_{\mu\nu}$, the scalars $g_{ab}$, $C_{abcd}$ and $\varphi$, and the
twelve $U(1)$ gauge bosons $B_{a\mu}$ and $C_{a\mu}$.  (Here $C$
denotes a RR potential, and $B$ the NS potential).  It is also
consistent with the orientifold projection to turn on internal NS and
RR three-form fluxes.  However, note that these fluxes are discrete by
Dirac quantization, and non-dynamical in the massless sector, since
the corresponding zero-modes are projected out.

In the absence of flux, this orientifold describes the same theory as
Type I on $T^6$, via T-duality in all six torus directions.  The
sixteen D9-branes of $SO(32)$ in Type I become sixteen D3-branes after
T-duality.  Also, the charge and tension of the D3-branes is cancelled
by $2^6$ O3-planes located at the fixed points of the $Z_2$.  So, the
low energy effective field theory is the same ${\cal N}=4$ $SO(32)$
super-Yang-Mills, coupled to ${\cal N}=4$ supergravity.

Once we replace some of the D3-branes with fluxes, this story is
modified.  The fluxes generate a potential for the scalars, and
correspond to turning on charges that couple the scalars to the twelve
$U(1)$ gauge fields.  The result is a superhiggs mechanism in which
many of the scalars get massive or are eaten by massive vectors,
breaking ${\cal N}=4$ to ${\cal N}<4$ supersymmetry.

As an example, consider the choice of flux~\cite{KST}
\begin{eqnarray}
\label{fluxes}
\frac{1}{(2\pi)^2\alpha'}F^{\rm RR}_{(3)} & = &
4dx^1\wedge dx^2\wedge dy^3 + 4dy^1\wedge dy^2\wedge dy^3,\\
\frac{1}{(2\pi)^2\alpha'}H^{\rm NS}_{(3)} & = &
4dx^1\wedge dx^2\wedge dx^3 + 4dy^1\wedge dy^2\wedge dx^3.
\end{eqnarray}
Let us parametrize the complex structure as $dz^i =
dx^i+\tau^i{}_jdy^j$, and normalize the holomorphic three-form so that
$\Omega = dz^1\wedge dz^2\wedge dz^3$.  Then, by wedging the
appropriate three-forms together, it is easy to show that
\begin{equation}
W = \int G\wedge\Omega \,\propto\, 1+ \bigl({\rm
cof}\,\tau\bigr)_3{}^3 +\varphi\Bigl(\det\tau + \tau^3{}_3\Bigr).
\end{equation}

For supersymmetric vacua, the equations of motion are that $D_I W =
\partial_I W + (\partial_I{\cal K}) W = 0$, where ${\cal
K}(\varphi^I,\bar{\varphi}^{\bar{I}})$ is the K\"ahler potential on
moduli space.  Since the superpotential is independent of K\"ahler
moduli, this simplifies to $W= \partial_\varphi W = \partial_\tau W =
0$, which through a small amount of algebra can be shown to imply that
\begin{equation}
\label{cpxstrmod}
\varphi\,\tau^3{}_3 = -1,
\quad \tau^1{}_1\,\tau^2{}_2-\tau^1{}_2\,\tau^2{}_1 = -1.
\end{equation}
So, the moduli space of complex structure is complex four-dimensional,
and can be parametrized by, say $\tau^1{}_1$, $\tau^2{}_2$,
$\tau^3{}_3$, and $\tau^1{}_2$.  In Eq.~(\ref{fluxes}), we expressed
the flux as a linear combination of integral three-forms with integer
coefficients, as required Dirac quantization.  Using
Eq.~(\ref{cpxstrmod}), we can also write the flux in terms of
holomorphic and antiholomorphic forms.  Restricting to $\tau^i{}_j$
diagonal and imaginary for simplicity, we find that
\begin{equation}
\label{holoflux}
G \propto dz^1\wedge d{\bar z}^{\bar2}\wedge dz^3
+ d{\bar z}^{\bar1}\wedge dz^2\wedge dz^3.
\end{equation}
This makes it clear that if Eq.~(\ref{cpxstrmod}) is satisfied then
the complex flux is indeed of type (2,1).  In addition, it is easy to
show that the primitivity condition is satisfied on the appropriate
subspace of K\"ahler moduli.

As a final remark, note that if we replace $z^1$ and $z^2$ by their
complex conjugates, then the flux~(\ref{holoflux}) is still (2,1) and
primitive.  In other words, there are two inequivalent complex
structures in which the conditions for ${\cal N}=1$ supersymmetry are
satisfied.  This implies that the solution is actually ${\cal N}=2$
supersymmetric, and shows how to engineer solutions with anywhere from
${\cal N}=0$ supersymmetry (when there is {\it no\/} solution to $D_I
W=0$) to ${\cal N} = 3$ supersymmetry (when the solution permits three
independent complex structures).  

For a more complete discussion, including large classes of ${\cal
N}=1$ solutions, we refer the reader to the work \cite{KST} on which
this review is based, and the references contained therein.


\begin{acknowledgments}
I would like to thank S.~Kachru and S.~P.~Trivedi for the
collaboration on which these proceedings are based.  In addition, I
would like to thank the organizers of the Carg\`ese 2002 ASI, as well
as M.~Berg, M.~Haack, and A.~Micu for useful discussions during the
course of the school.  This work was supported in part by the DOE
under contracts DE-AC03-76SF00515 and DE-FG03-92-ER40701.
\end{acknowledgments}

\end{document}